\documentclass[utf8]{frontiersSCNS} 

\usepackage[onehalfspacing]{setspace}

\usepackage{color}

\def\keyFont{\fontsize{8}{11}\helveticabold }
\def\firstAuthorLast{Bisogni {et~al.}} 
\def\Authors{Susanna Bisogni\,$^{1,2,*}$, Guido Risaliti\,$^{3,2}$ and Elisabeta Lusso\,$^{4, 2}$}
\def\Address{$^{1}$Harvard-Smithsonian Center for Astrophysics, Cambridge, MA, USA \\
$^{2}$INAF, Osservatorio Astrofisico di Arcetri, Firenze, Italy\\
$^{3}$Universit\`a degli Studi di Firenze, Dipartimento di Fisica e Astronomia, Sesto Fiorentino, Firenze, Italy \\
$^{4}$Centre for Extragalactic Astronomy, Department of Physics, Durham University, Durham, UK   }
\def\corrAuthor{Susanna Bisogni}

\def\corrEmail{susanna.bisogni@cfa.harvard.edu}

\begin{document}
\onecolumn
\firstpage{1}

\title[Quasars as Standard Candles]{A Hubble Diagram for Quasars}

\author[\firstAuthorLast ]{\Authors} 
\address{} 
\correspondance{} 

\extraAuth{}

\maketitle

\begin{abstract}

\section{}
The cosmological model is at present not tested between the redshift of the farthest observed supernovae ($z\sim 1.4$) and that of the Cosmic Microwave Background ($z \sim 1,100$). Here we introduce a new method to measure the cosmological parameters: we show that quasars can be used as ``standard candles'' by employing the non-linear relation between their intrinsic UV and X-ray emission as an absolute distance indicator. We built a sample of $\sim 1,900$ quasars with available UV and X-ray observations, and produced a Hubble Diagram up to $z \sim 5$. The analysis of the quasar Hubble Diagram, when used in combination with supernovae, provides robust constraints on the matter and energy content in the cosmos. The application of this method to forthcoming, larger quasar samples, will also provide tight constraints on the dark energy equation of state and its possible evolution with time.

\tiny
 \keyFont{ \section{Keywords:} cosmology: distance scale -- cosmological parameters -- observations, galaxies: active, quasars: general, X-ray:general, ultraviolet: general }
\end{abstract}

\section{Introduction}

Quasars are among the brightest sources in the universe, by now observable up to redshift $z \sim 7$ \citep{Mortlock2011}. For this reason they have always been considered as potential candidates for extending the distance ladder in a redshift range well beyond the limit imposed by the supernovae ($z\sim 1.4$).
Quasars, however,  are known to be extremely variable, anisotropic sources and characterised by a wide range in luminosity. 
Unlike supernovae, an ``easy'' connection between a spectral (or time-dependent) feature and the luminosity is not available. In short, the use of quasars as standard candles is not obvious.
The fundamental requirement needed to employ these sources for a cosmological purpose is being able to measure a ``standard luminosity'' from which infer the distance. 

Several attempts have been made in this sense, using different relations involving quasars luminosity, such as the \emph{Baldwin effect} \citep{Baldwin1977, Korista1998}, the Broad Line Region radius - luminosity relation \citep{Watson2011, KilerciEser2015}, the wavelength-dependent time delays in the emission variability of the accretion disc \citep{Collier1999}, the distribution of the linewidths as a function of the inclination of the BLR and of the quasar luminosity function \citep{RudgeRaine1999}, the properties of highly accreting quasars \citep{Wang2013, MarzianiSulentic2014}, the relation between the mass of the SuperMassive Black Hole and the X-ray variability \citep{LaFranca2014} among the others, or invoking the use of classic geometrical arguments, such as the measurement of the apparent size of the Broad Line Region \citep{ElvisKarovska2002}, whose intrinsic dimension can be inferred from reverberation mapping monitoring \citep{BlandfordMcKee1982, Peterson1993}.
Most of these methods suffer from the high scatter in the observed relation or are limited by a poor statistics - generally imposed by the long times required by the observations.
In order to make cosmology with quasars we need to 1) have a measurement of their distance with a high precision, i.e we need a luminosity-related relation with the smallest possible scatter; 2) make the most out of their large statistics, i.e. being able to apply the method to a large number of sources.

We recently proposed the relation between the luminosities in the X-rays ($2$keV) and UV ($2500$\AA) bands \citep{Tananbaum1979, Zamorani1981,Steffen2006, Lusso2010, Young2010, LussoRisaliti2016, LussoRisaliti2017} as a method to estimate quasars distances \citep{RisalitiLusso2015}.
The method relies on the simple fact that the $L_{X}-L_{UV}$ relation is not linear. If we express it in terms of the fluxes, it becomes a function of the slope, the intercept, the intrinsic dispersion and the luminosity distance, i.e. a function of the cosmological parameters $\Omega_{M}$ and $\Omega_{\Lambda}$.
We can then use the fit to the relation or, equivalently, we can build the Hubble diagram for quasars, to infer information on the cosmological parameters.
The obvious advantage of this method is that it only requires a measure of the two fluxes to be employed, therefore avoiding long-monitoring programs and allowing its application on a large number of sources up to high redshifts.

\section{Sample}

In \cite{RisalitiLusso2015} we presented the first Hubble diagram for a sample of $\sim800$ quasars for which $2$keV and $2500$\AA~ measures were available from the literature.
We extended the Hubble diagram in a redshift range that was never explored before by any other cosmological probe (up to $z\sim6$) and verified the excellent agreement between the distance moduli inferred with supernovae and those inferred with quasars in the common redshift range ($z=0.01-1.4$).

The reliability and effectiveness of the method strongly depend, respectively, on the non-evolution of the relation with reshift and the dispersion in the relation, that, as already mentioned, directly affects the precision in distance estimates.
This first sample, although properly treated to suit cosmological studies, can be improved in terms of homogeneity, quality and statistics of the data. 
In order to do that we selected a new sample of $\sim8000$ objects for which X-rays and UV observations are available, crossmatching the SDSS Data Release 7 \citep{Shen2011} and Data Release 12 \citep{Paris2017} catalogues with the 3XMM-DR5 \citep{Rosen2016} catalogue. 
The much higher statistics of this sample with respect to the previous one allows us to apply stronger quality cuts in order to significantly decrease the dispersion in the $L_{X}-L_{UV}$ relation and eventually to verify to a much higher extent the non-evolution of the relation with redshift. 

\subsection{The dispersion in the $L_{X}-L_{UV}$ relation}

The $L_{X}-L_{UV}$ relation has been known to be characterised by a high dispersion ($\sim 0.3-0.4$ dex) that has deterred from using it as an absolute distance indicator.
The observed dispersion, however, is the result of two distinct contributions: an intrinsic scatter in the relation, related to the still unknown physics involved, and an additional scatter due to observational issues.
In \cite{LussoRisaliti2016} and \cite{LussoRisaliti2017} we demonstrated that a large part of the observed dispersion is ascribable to the latter contribution and, through the analysis of a smaller sample of high quality data and of sources with multiple observations available, that the magnitude of the \textit{instrinsic dispersion} can be attested to be $<0.20$ dex.
Removing the part of the observed dispersion related to observational issues makes the relation much tighter and our distance estimates much more precise. Significant efforts has then been made in this direction, once the major contributions to this ``observational'' scatter has been recognised in the following issues:

\begin{itemize}
\item[1.] uncertainties in the measurement of the ($2$ keV) X-ray flux, which are, on average, a factor of 2 (unlike UV measurements that can be constrained with uncertainties lower than 10\%),
\item[2.] absorption in the spectrum in the UV and in the X-ray wavelength ranges (with the effect on the UV band being more severe and also less recognisable with respect to the one on the X-ray spectrum),
\item[3.] variability of the source and non-simultaneity of the observation in the UV and X-ray bands,
\item[4.] inclination effects affecting the intrinsic emission of the accretion disc, 
\item[5.] selection effects due to the flux limit in the surveys (the Eddington Bias, i.e. sources with an average flux below the detection threshold can be detected only while on a positive fluctuation).
\end{itemize}

\begin{figure}[h!]
\begin{center}
\includegraphics[width=13cm]{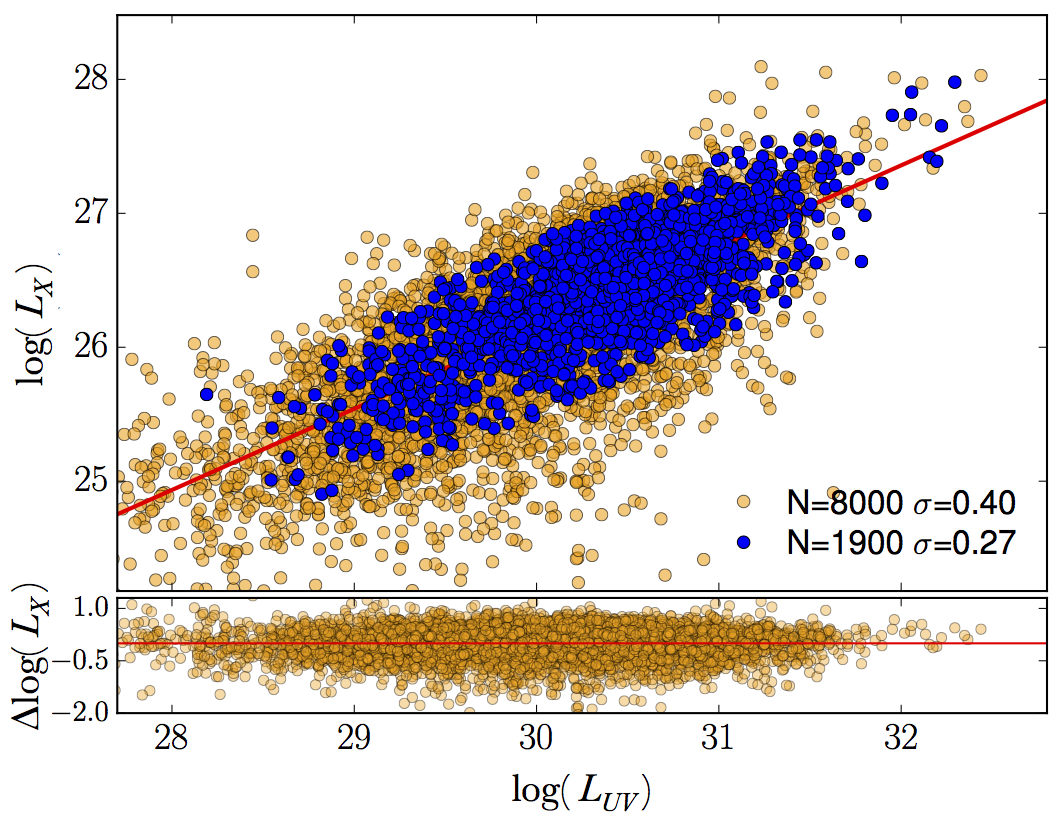}
\end{center}
\caption{Comparison between the relation obtained for the whole sample of quasars for which UV and X-ray observations are available ($\sim 8000$) and for the sample cleaned by the observational issues that contribute to the observed dispersion ($\sim 1900$). After the quality cuts are applied, the observed dispersion is decreased by $\sim 0.3$ dex.}\label{fig1}
\end{figure}

By filtering the sample according to the quality of the X-ray measurements, the amount of intrinsic absorption in the source and the systematic effects due to the Eddington bias, we obtain a ``clean'' sample with a scatter in the relation reduced by $\sim 0.3$ dex\footnote{Since from the initial sample selection, jetted quasars and Broad Absorption Line quasars (BAL) were excluded. These sources are known, respectively, to have an additional contributions to the X-ray emission (with respect to the one coming from the hot corona) and to be heavily obscured.} (Fig. \ref{fig1}). These criteria select a sample of $\sim 1900$ objects out of the initial $\sim 8000$ with a dispersion of $\sim 0.27$ dex.
Considering that the dispersion in the $L_{X}-L_{UV}$ relation is propagated to the distance modulus and that the dispersion in the Hubble diagram for supernovae 1A is $\sim 0.07$ at $z \sim 1$, this means that $\sim 15$ quasars can provide the same cosmological information as one supernova at this redshift.

\subsection{The non-evolution of the $L_{X}-L_{UV}$ relation with redshift}

For the relation to be used as a reliable method for distance estimates, we need to check its non-evolution with the redshift. We repeated the analysis performed in \cite{RisalitiLusso2015} on the new, larger sample: we divided the sample in redshift bins and evaluated the relation within each one of them.
The size of the redshift bin must fullfill two requirements: it has to be small enough so that differences in the luminosity distance within the same bin are negligible with respect to the intrinsic dispersion in the relation (the observed dispersion in each redshift bin is $\sim 0.25$ dex on average), but large enough so that the number of sources within each bin makes this analysis meaningful.
Moreover, if the differences in the luminosity distances are negligible, we can consider the relation between fluxes instead of luminosities, making this check independent from the assumed cosmological model.

The two requirements are met when the width of the redshift  bin is $\Delta \mathrm{log} z = 0.08$. The whole redshift range spanned by quasars (with the exclusion of the range $z \cong 0.01-0.3$, for which a significant number of objects is not available) yields twelve redshift bins.


For each subsample in the twelve redshift bins, we fitted the X-ray to UV {\em fluxes} with the same log-linear relation adopted for the luminosities of the whole sample. The results of these fits for the slope parameter are shown in Fig. 2. All the fits are consistent with $\alpha=0.6$, with an average $<\alpha>$=0.56$\pm$0.08.

\begin{figure}[h!]
\begin{center}
\includegraphics[width=17cm]{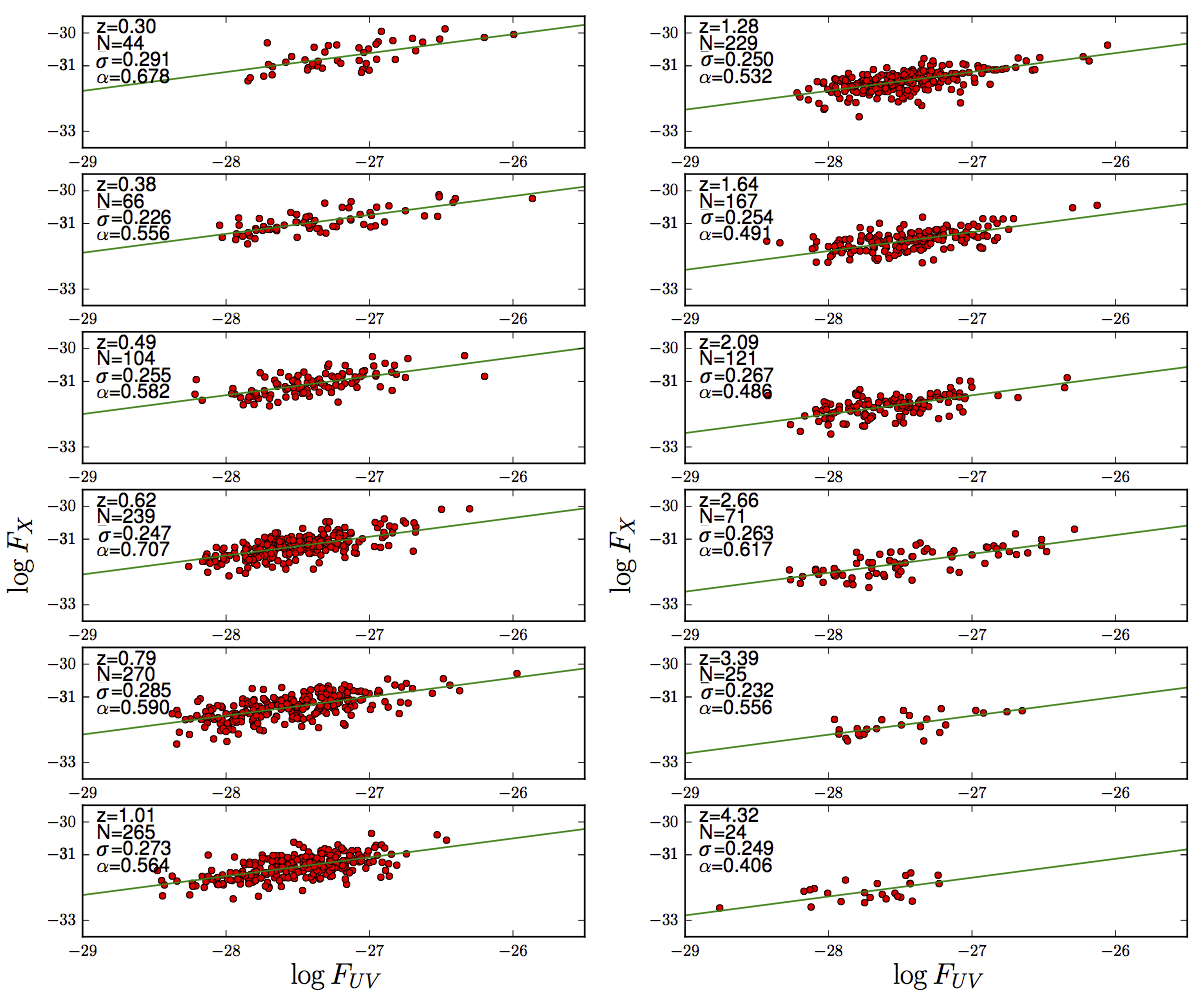}
\end{center}
\caption{Fits of the $F_{X}-F_{UV}$ in narrow redshift intervals. The distance differences among the objects in the same bin are small with respect to the dispersion of the relation, so we can use fluxes as proxies of the luminosities. In this way, the analysis is independent from the choice of a cosmological model. In each panel we report the median redshift, the number of objects $N$, the dispersion $\sigma$, and the best fit slope $\alpha$ of the relation. }\label{fig2}
\end{figure}

\section{The Hubble Diagram for quasars}

\begin{figure}[h!]
\begin{center}
\includegraphics[width=17cm]{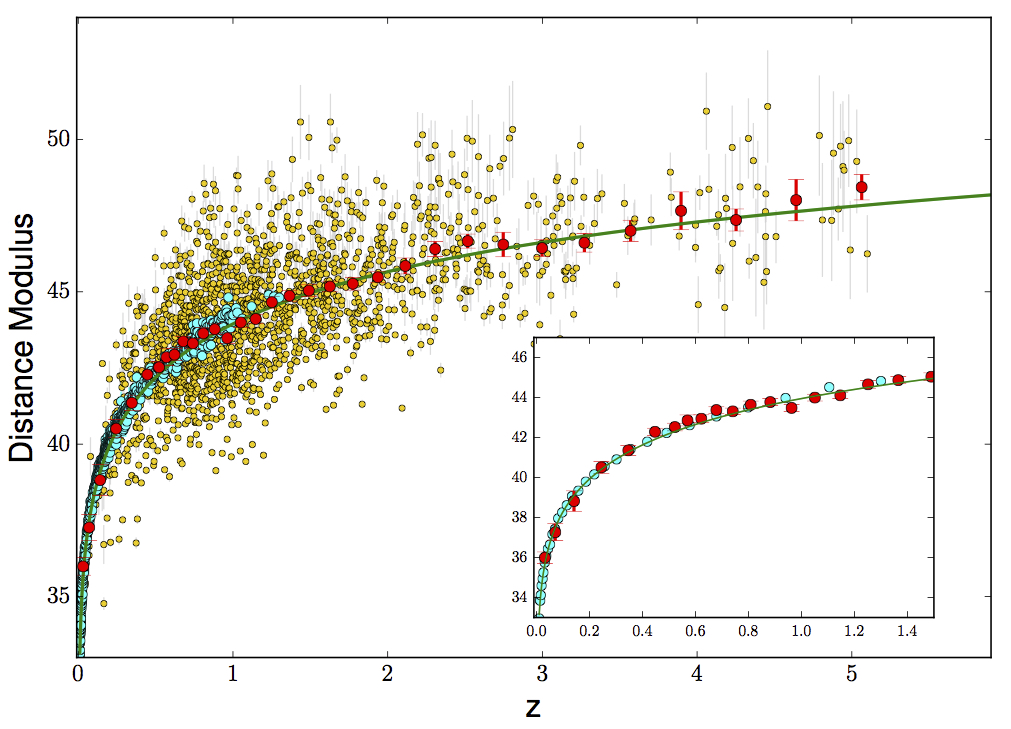}
\end{center}
\caption{Hubble Diagram of quasars obtained with our ``clean'' sample and supernovae 1A. The orange points are single measurements for quasars, while the red points are quasar averages in small redshift bins. Type 1A supernovae are also plotted with cyan points \citep[JLA sample,][]{Betoule2014}. The inset plot shows a zoom of the same diagram in the redshift range where supernova 1A and quasars overlap. In this case both red and cyan points are averages in small redshift bins for quasars and supernovae 1A respectively.}\label{fig3}
\end{figure}

The $L_{X}-L_{UV}$ relation has been used to estimate quasar distances and build a Hubble diagram of quasars. While slope of the relation can be obtained in a cosmology-independent way as described in the previous Section, its absolute calibration requires the comparison with other standard candles (analogously to the calibration of SN1a based on Cepheid stars). We obtained such a calibration by requiring an overlap of the Hubble diagram of SN1a and quasars in the overlapping redshift range, $z \sim 0.2-1.4$. The result is shown in Fig.\ref{fig3}. The main interesting aspects of this work are the following:\\
1. The Hubble diagram of quasars perfectly overlap with the one of SN1a in the redshift range $z=0.2-1.4$: the same cosmological model fits both the supernovae and quasars data, with no significant residuals beyond the expected Gaussian fluctuations (see inset of Fig. 3).  While the absolute calibration of quasars has been chosen in order to have such an agreement, the shape of the diagram has no free parameter once the slope of the relation has been fixed. Therefore, the match over the whole common redshift range is a strong confirmation of the reliability of our method.\\
2. There are a few hundred quasars at redshifts higher than $z=1.4$, where no supernova has been observed. The Hubble diagram of these objects is a powerful check of the cosmological model, tracing the evolution of the Universe in the poorly investigated $z=1.4-6$ redshift interval, the only other cosmological probes testing the expansion of the Universe above $z=1.4$ being Gamma Ray Bursts \citep{Ghirlanda2006} and Ly-$\alpha$ BAO at $z\sim2.4$ \citep{duMas2017}.\\ 
3. The average dispersion in the $\log(D_L)$-redshift relation is $\sigma\sim0.27$~dex. This is a large improvement with respect to our first work based on literature samples ($\sigma\sim0.35$) and also on our first Hubble diagram based on SDSS quasars \citep[$\sigma\sim0.30$,][]{Risaliti2017}. This is due to our on-going refinement of both UV and X-ray flux measurements and sample selection. Our primary goal in the early phases of this new branch of observational cosmology is to obtain a clean sample, where biases and systematic effects are greatly reduced (a complete discussion of this point will be presented in a forthcoming paper, together with the results of the fits of the Hubble Diagram with several cosmological models). This comes at the expense of sample statistics: the final sample in Fig.~\ref{fig3} consists of ``only'' $1900$ objects, out of a parent sample of about $8000$ quasars. The main rejection criteria are: X-ray absorption, optical-UV reddening, poor quality of X-ray observations. In the future a more detailed spectral analysis of both the X-ray and optical/UV spectra will allow us to recover a large fraction of the rejected quasars, greatly enhancing the power of this method in constraining the cosmological parameters. 

\section*{Conflict of Interest Statement}

The authors declare that the research was conducted in the absence of any commercial or financial relationships that could be construed as a potential conflict of interest.

\section*{Funding}
Support for this work was provided by the National Aeronautics and Space Administration through Chandra Award Number AR7-18013 X issued by the Chandra X-ray Observatory Center, which is operated by the Smithsonian Astrophysical Observatory for and on behalf of the National Aeronautics Space Administration under contract NAS8-03060. E.L. is supported by a European Union COFUND/Durham Junior Research Fellowship (under EU grant agreement no. 609412).

\bibliographystyle{frontiersinSCNS_ENG_HUMS} 
\bibliography{QasSC}

\begin{thebibliography}{28}
\providecommand{\natexlab}[1]{#1}
\expandafter\ifx\csname urlstyle\endcsname\relax
  \providecommand{\doi}[1]{doi:\discretionary{}{}{}#1}\else
  \providecommand{\doi}{doi:\discretionary{}{}{}\begingroup
  \urlstyle{rm}\Url}\fi
\providecommand{\selectlanguage}[1]{\relax}
\providecommand{\bibAnnoteFile}[1]{%
  \IfFileExists{#1}{\begin{quotation}\noindent\textsc{Key:} #1\\
  \textsc{Annotation:}\ \input{#1}\end{quotation}}{}}
\providecommand{\bibAnnote}[2]{%
  \begin{quotation}\noindent\textsc{Key:} #1\\
  \textsc{Annotation:}\ #2\end{quotation}}

\bibitem[{{Baldwin}(1977)}]{Baldwin1977}
{Baldwin}, J.~A. (1977).
\newblock {Luminosity Indicators in the Spectra of Quasi-Stellar Objects}.
\newblock \emph{\apj} 214, 679--684.
\newblock \doi{10.1086/155294}
\bibAnnoteFile{Baldwin1977}

\bibitem[{{Betoule} et~al.(2014){Betoule}, {Kessler}, {Guy}, {Mosher},
  {Hardin}, {Biswas} et~al.}]{Betoule2014}
{Betoule}, M., {Kessler}, R., {Guy}, J., {Mosher}, J., {Hardin}, D., {Biswas},
  R., et~al. (2014).
\newblock {Improved cosmological constraints from a joint analysis of the
  SDSS-II and SNLS supernova samples}.
\newblock \emph{\aap} 568, A22.
\newblock \doi{10.1051/0004-6361/201423413}
\bibAnnoteFile{Betoule2014}

\bibitem[{{Blandford} and {McKee}(1982)}]{BlandfordMcKee1982}
{Blandford}, R.~D. and {McKee}, C.~F. (1982).
\newblock {Reverberation mapping of the emission line regions of Seyfert
  galaxies and quasars}.
\newblock \emph{\apj} 255, 419--439.
\newblock \doi{10.1086/159843}
\bibAnnoteFile{BlandfordMcKee1982}

\bibitem[{{Collier} et~al.(1999){Collier}, {Horne}, {Wanders}, and
  {Peterson}}]{Collier1999}
{Collier}, S., {Horne}, K., {Wanders}, I., and {Peterson}, B.~M. (1999).
\newblock {A new direct method for measuring the Hubble constant from
  reverberating accretion discs in active galaxies}.
\newblock \emph{\mnras} 302, L24--L28.
\newblock \doi{10.1046/j.1365-8711.1999.02250.x}
\bibAnnoteFile{Collier1999}

\bibitem[{{du Mas des Bourboux} et~al.(2017){du Mas des Bourboux}, {Le Goff},
  {Blomqvist}, {Busca}, {Guy}, {Rich} et~al.}]{duMas2017}
{du Mas des Bourboux}, H., {Le Goff}, J.-M., {Blomqvist}, M., {Busca}, N.~G.,
  {Guy}, J., {Rich}, J., et~al. (2017).
\newblock {Baryon acoustic oscillations from the complete SDSS-III
  Ly$\alpha$-quasar cross-correlation function at $z=2.4$}.
\newblock \emph{ArXiv e-prints}
\bibAnnoteFile{duMas2017}

\bibitem[{{Elvis} and {Karovska}(2002)}]{ElvisKarovska2002}
{Elvis}, M. and {Karovska}, M. (2002).
\newblock {Quasar Parallax: A Method for Determining Direct Geometrical
  Distances to Quasars}.
\newblock \emph{\apjl} 581, L67--L70.
\newblock \doi{10.1086/346015}
\bibAnnoteFile{ElvisKarovska2002}

\bibitem[{{Ghirlanda} et~al.(2006){Ghirlanda}, {Ghisellini}, and
  {Firmani}}]{Ghirlanda2006}
{Ghirlanda}, G., {Ghisellini}, G., and {Firmani}, C. (2006).
\newblock {Gamma-ray bursts as standard candles to constrain the cosmological
  parameters}.
\newblock \emph{New Journal of Physics} 8, 123.
\newblock \doi{10.1088/1367-2630/8/7/123}
\bibAnnoteFile{Ghirlanda2006}

\bibitem[{{Kilerci Eser} et~al.(2015){Kilerci Eser}, {Vestergaard}, {Peterson},
  {Denney}, and {Bentz}}]{KilerciEser2015}
{Kilerci Eser}, E., {Vestergaard}, M., {Peterson}, B.~M., {Denney}, K.~D., and
  {Bentz}, M.~C. (2015).
\newblock {On the Scatter in the Radius-Luminosity Relationship for Active
  Galactic Nuclei}.
\newblock \emph{\apj} 801, 8.
\newblock \doi{10.1088/0004-637X/801/1/8}
\bibAnnoteFile{KilerciEser2015}

\bibitem[{{Korista} et~al.(1998){Korista}, {Baldwin}, and
  {Ferland}}]{Korista1998}
{Korista}, K., {Baldwin}, J., and {Ferland}, G. (1998).
\newblock {Quasars as Cosmological Probes: The Ionizing Continuum, Gas
  Metallicity, and the W$_{\lambda}$-L Relation}.
\newblock \emph{\apj} 507, 24--30.
\newblock \doi{10.1086/306321}
\bibAnnoteFile{Korista1998}

\bibitem[{{La Franca} et~al.(2014){La Franca}, {Bianchi}, {Ponti}, {Branchini},
  and {Matt}}]{LaFranca2014}
{La Franca}, F., {Bianchi}, S., {Ponti}, G., {Branchini}, E., and {Matt}, G.
  (2014).
\newblock {A New Cosmological Distance Measure Using Active Galactic Nucleus
  X-Ray Variability}.
\newblock \emph{\apjl} 787, L12.
\newblock \doi{10.1088/2041-8205/787/1/L12}
\bibAnnoteFile{LaFranca2014}

\bibitem[{{Lusso} et~al.(2010){Lusso}, {Comastri}, {Vignali}, {Zamorani},
  {Brusa}, {Gilli} et~al.}]{Lusso2010}
{Lusso}, E., {Comastri}, A., {Vignali}, C., {Zamorani}, G., {Brusa}, M.,
  {Gilli}, R., et~al. (2010).
\newblock {The X-ray to optical-UV luminosity ratio of X-ray selected type 1
  AGN in XMM-COSMOS}.
\newblock \emph{\aap} 512, A34.
\newblock \doi{10.1051/0004-6361/200913298}
\bibAnnoteFile{Lusso2010}

\bibitem[{{Lusso} and {Risaliti}(2016)}]{LussoRisaliti2016}
{Lusso}, E. and {Risaliti}, G. (2016).
\newblock {The Tight Relation between X-Ray and Ultraviolet Luminosity of
  Quasars}.
\newblock \emph{\apj} 819, 154.
\newblock \doi{10.3847/0004-637X/819/2/154}
\bibAnnoteFile{LussoRisaliti2016}

\bibitem[{{Lusso} and {Risaliti}(2017)}]{LussoRisaliti2017}
{Lusso}, E. and {Risaliti}, G. (2017).
\newblock {Quasars as standard candles. I. The physical relation between disc
  and coronal emission}.
\newblock \emph{\aap} 602, A79.
\newblock \doi{10.1051/0004-6361/201630079}
\bibAnnoteFile{LussoRisaliti2017}

\bibitem[{{Marziani} and {Sulentic}(2014)}]{MarzianiSulentic2014}
{Marziani}, P. and {Sulentic}, J.~W. (2014).
\newblock {Highly accreting quasars: sample definition and possible
  cosmological implications}.
\newblock \emph{\mnras} 442, 1211--1229.
\newblock \doi{10.1093/mnras/stu951}
\bibAnnoteFile{MarzianiSulentic2014}

\bibitem[{{Mortlock} et~al.(2011){Mortlock}, {Warren}, {Venemans}, {Patel},
  {Hewett}, {McMahon} et~al.}]{Mortlock2011}
{Mortlock}, D.~J., {Warren}, S.~J., {Venemans}, B.~P., {Patel}, M., {Hewett},
  P.~C., {McMahon}, R.~G., et~al. (2011).
\newblock {A luminous quasar at a redshift of z = 7.085}.
\newblock \emph{\nat} 474, 616--619.
\newblock \doi{10.1038/nature10159}
\bibAnnoteFile{Mortlock2011}

\bibitem[{{P{\^a}ris} et~al.(2017){P{\^a}ris}, {Petitjean}, {Ross}, {Myers},
  {Aubourg}, {Streblyanska} et~al.}]{Paris2017}
{P{\^a}ris}, I., {Petitjean}, P., {Ross}, N.~P., {Myers}, A.~D., {Aubourg},
  {\'E}., {Streblyanska}, A., et~al. (2017).
\newblock {The Sloan Digital Sky Survey Quasar Catalog: Twelfth data release}.
\newblock \emph{\aap} 597, A79.
\newblock \doi{10.1051/0004-6361/201527999}
\bibAnnoteFile{Paris2017}

\bibitem[{{Peterson}(1993)}]{Peterson1993}
{Peterson}, B.~M. (1993).
\newblock {Reverberation mapping of active galactic nuclei}.
\newblock \emph{\pasp} 105, 247--268.
\newblock \doi{10.1086/133140}
\bibAnnoteFile{Peterson1993}

\bibitem[{{Risaliti} and {Lusso}(2015)}]{RisalitiLusso2015}
{Risaliti}, G. and {Lusso}, E. (2015).
\newblock {A Hubble Diagram for Quasars}.
\newblock \emph{\apj} 815, 33.
\newblock \doi{10.1088/0004-637X/815/1/33}
\bibAnnoteFile{RisalitiLusso2015}

\bibitem[{{Risaliti} and {Lusso}(2017)}]{Risaliti2017}
{Risaliti}, G. and {Lusso}, E. (2017).
\newblock {Cosmology with AGN: can we use quasars as standard candles?}
\newblock \emph{Astronomische Nachrichten} 338, 329--333.
\newblock \doi{10.1002/asna.201713351}
\bibAnnoteFile{Risaliti2017}

\bibitem[{{Rosen} et~al.(2016){Rosen}, {Webb}, {Watson}, {Ballet}, {Barret},
  {Braito} et~al.}]{Rosen2016}
{Rosen}, S.~R., {Webb}, N.~A., {Watson}, M.~G., {Ballet}, J., {Barret}, D.,
  {Braito}, V., et~al. (2016).
\newblock {The XMM-Newton serendipitous survey. VII. The third XMM-Newton
  serendipitous source catalogue}.
\newblock \emph{\aap} 590, A1.
\newblock \doi{10.1051/0004-6361/201526416}
\bibAnnoteFile{Rosen2016}

\bibitem[{{Rudge} and {Raine}(1999)}]{RudgeRaine1999}
{Rudge}, C.~M. and {Raine}, D.~J. (1999).
\newblock {Determining the cosmological parameters from the linewidths of
  active galaxies}.
\newblock \emph{\mnras} 308, 1150--1158.
\newblock \doi{10.1046/j.1365-8711.1999.02802.x}
\bibAnnoteFile{RudgeRaine1999}

\bibitem[{{Shen} et~al.(2011){Shen}, {Richards}, {Strauss}, {Hall},
  {Schneider}, {Snedden} et~al.}]{Shen2011}
{Shen}, Y., {Richards}, G.~T., {Strauss}, M.~A., {Hall}, P.~B., {Schneider},
  D.~P., {Snedden}, S., et~al. (2011).
\newblock {A Catalog of Quasar Properties from Sloan Digital Sky Survey Data
  Release 7}.
\newblock \emph{\apjs} 194, 45.
\newblock \doi{10.1088/0067-0049/194/2/45}
\bibAnnoteFile{Shen2011}

\bibitem[{{Steffen} et~al.(2006){Steffen}, {Strateva}, {Brandt}, {Alexander},
  {Koekemoer}, {Lehmer} et~al.}]{Steffen2006}
{Steffen}, A.~T., {Strateva}, I., {Brandt}, W.~N., {Alexander}, D.~M.,
  {Koekemoer}, A.~M., {Lehmer}, B.~D., et~al. (2006).
\newblock {The X-Ray-to-Optical Properties of Optically Selected Active
  Galaxies over Wide Luminosity and Redshift Ranges}.
\newblock \emph{\aj} 131, 2826--2842.
\newblock \doi{10.1086/503627}
\bibAnnoteFile{Steffen2006}

\bibitem[{{Tananbaum} et~al.(1979){Tananbaum}, {Avni}, {Branduardi}, {Elvis},
  {Fabbiano}, {Feigelson} et~al.}]{Tananbaum1979}
{Tananbaum}, H., {Avni}, Y., {Branduardi}, G., {Elvis}, M., {Fabbiano}, G.,
  {Feigelson}, E., et~al. (1979).
\newblock {X-ray studies of quasars with the Einstein Observatory}.
\newblock \emph{\apjl} 234, L9--L13.
\newblock \doi{10.1086/183100}
\bibAnnoteFile{Tananbaum1979}

\bibitem[{{Wang} et~al.(2013){Wang}, {Du}, {Valls-Gabaud}, {Hu}, and
  {Netzer}}]{Wang2013}
{Wang}, J.-M., {Du}, P., {Valls-Gabaud}, D., {Hu}, C., and {Netzer}, H. (2013).
\newblock {Super-Eddington Accreting Massive Black Holes as Long-Lived
  Cosmological Standards}.
\newblock \emph{Physical Review Letters} 110, 081301.
\newblock \doi{10.1103/PhysRevLett.110.081301}
\bibAnnoteFile{Wang2013}

\bibitem[{{Watson} et~al.(2011){Watson}, {Denney}, {Vestergaard}, and
  {Davis}}]{Watson2011}
{Watson}, D., {Denney}, K.~D., {Vestergaard}, M., and {Davis}, T.~M. (2011).
\newblock {A New Cosmological Distance Measure Using Active Galactic Nuclei}.
\newblock \emph{\apjl} 740, L49.
\newblock \doi{10.1088/2041-8205/740/2/L49}
\bibAnnoteFile{Watson2011}

\bibitem[{{Young} et~al.(2010){Young}, {Elvis}, and {Risaliti}}]{Young2010}
{Young}, M., {Elvis}, M., and {Risaliti}, G. (2010).
\newblock {The X-ray Energy Dependence of the Relation Between Optical and
  X-ray Emission in Quasars}.
\newblock \emph{\apj} 708, 1388--1397.
\newblock \doi{10.1088/0004-637X/708/2/1388}
\bibAnnoteFile{Young2010}

\bibitem[{{Zamorani} et~al.(1981){Zamorani}, {Henry}, {Maccacaro}, {Tananbaum},
  {Soltan}, {Avni} et~al.}]{Zamorani1981}
{Zamorani}, G., {Henry}, J.~P., {Maccacaro}, T., {Tananbaum}, H., {Soltan}, A.,
  {Avni}, Y., et~al. (1981).
\newblock {X-ray studies of quasars with the Einstein Observatory. II}.
\newblock \emph{\apj} 245, 357--374.
\newblock \doi{10.1086/158815}
\bibAnnoteFile{Zamorani1981}

\end{thebibliography}

\end{document}